\documentclass[11pt]{article}

\usepackage[T1]{fontenc}
\usepackage[utf8]{inputenc}
\usepackage[margin=1.15in]{geometry}
\usepackage{times}
\usepackage{microtype}
\usepackage{amsmath,amssymb}
\usepackage{booktabs}
\usepackage{makecell}
\usepackage{graphicx}
\usepackage{listings}
\usepackage[dvipsnames]{xcolor}
\usepackage{tikz}
\usetikzlibrary{arrows.meta,positioning,fit,backgrounds,calc}
\usepackage{caption}
\usepackage[numbers,sort&compress]{natbib}
\usepackage[colorlinks=true,allcolors=NavyBlue]{hyperref}
\usepackage{cleveref}

\newcommand{\code}[1]{\nolinkurl{#1}}

\definecolor{codebg}{HTML}{F5F4F1}
\definecolor{codekw}{HTML}{7A3E2E}
\newcommand{\CleanAcc}{0.860}
\newcommand{\CleanN}{500}
\newcommand{\PoisonN}{120}
\newcommand{\PoisonTotal}{360}
\newcommand{\PoisonFrac}{1.2\%}
\newcommand{\AccClean}{0.850}
\newcommand{\AccUndef}{0.300}
\newcommand{\AccShipped}{0.317}

\newcommand{\PShipped}{0.80}

\newcommand{\MarginShipped}{0.175}
\newcommand{\MarginCorrected}{0.544}
\newcommand{\PoisonGain}{0.32}
\newcommand{\FprAegis}{1.5\%}
\newcommand{\FprDeberta}{42.8\%}
\newcommand{\FprPromptGuard}{6.5\%}
\newcommand{\ScreenCaught}{0}
\newcommand{\MixBUFrac}{18.7\%}
\newcommand{\MixPoisonFrac}{1.18\%}
\newcommand{\MixOff}{0.3167}
\newcommand{\MixW}{0.7000}
\newcommand{\EvuOff}{0.8583}
\newcommand{\EvuW}{0.0417}
\newcommand{\AbsPenalty}{0.245}
\newcommand{\MaxSem}{0.45}
\newcommand{\AbstainRate}{96.2\%}
\newcommand{\EvalSpend}{11.22}
\newcommand{\StageTwoInjecAgent}{155}
\newcommand{\RecallDirect}{0.741}
\newcommand{\RecallIndirect}{0.832}
\newcommand{\RecallDirectCore}{0.144}

\title{\bfseries Utility Under Attack: Agent Memory Poisoning\\
and the Limits of Content Screening and Provenance Ranking}

\author{%
  Arulnidhi Karunanidhi\\
  \small Quantify Labs Ltd\\
  \small\texttt{arulnidhi@quantifylabs.ai}
}
\date{\today}

\begin{document}
\maketitle

\begin{abstract}
Persistent memory gives an attacker something a single request does not: a false statement accepted
once is retrieved into every future session that matches it. We measure what such an attack costs,
using the weakest form of it we could construct --- plainly-worded false assertions generated in a single pass,
carrying no instruction, no trigger, and no optimization against the retriever. At \PoisonFrac{} of
the corpus, this removes two-thirds of an agent memory's value on LongMemEval (accuracy \AccClean{}
to \AccUndef{}), and a four-stage write-time content screening pipeline --- one that reaches
\RecallIndirect{} recall on indirect prompt injection while flagging only \FprAegis{} of
trigger-word-laden benign text --- refuses \ScreenCaught{} of \PoisonTotal{} poisoned memories. We
argue this marks a boundary of content-only screening rather than a detector deficiency: distinguishing
a false assertion from a true one generally requires external grounding beyond the text being screened.

The defensive burden therefore falls on retrieval, where provenance-weighted ranking prefers content
from trusted channels. Two results follow, neither flattering to the system we build. The shipped
weight was statistically indistinguishable from no defense ($p{=}\PShipped{}$), for a reason
derivable in three lines from the scoring function. Raising it recovers utility, but by exclusion
rather than preference: its absolute penalty on untrusted content exceeds half the range available
to semantic similarity. We show this with two further corpora. Where untrusted content is
predominantly benign, the defense still works (\MixOff{} to \MixW{}); where the answer-bearing
evidence itself arrives untrusted, retrieval collapses to zero --- no evidence memory survives
ranking for any of 120 questions, and accuracy falls to \EvuW{}. In this measured similarity regime, we find no usable setting for the additive provenance term:
any weight sufficient against an attacker who can shape content is also sufficient to exclude untrusted
content categorically. We argue provenance belongs in retrieval as a
bounded occupancy constraint instead, and release the harnesses, corpora, and aggregate run reports.

\end{abstract}

\section{Introduction}
\label{sec:intro}

Agents that remember are agents that can be lied to durably. When an LLM application is stateless,
whatever an attacker achieves is bounded by a single request; when it writes to persistent memory,
a false statement accepted once is retrieved into every future session that matches it. The
security literature for LLM agents has concentrated on the first setting, where the threat is
instructions hidden in content and the defenses are detection, instruction hierarchies, and
constraints on what untrusted data may cause. This paper is about a threat that none of those
defenses is built to see, because it does not involve instructions at all.

We poisoned an established memory benchmark with false statements. Not optimized adversarial
strings, not hidden instructions, not embedding-space triggers --- ordinary sentences asserting
wrong facts, produced in a single generation pass with no iteration of any kind. At \PoisonFrac{}
of the corpus, that attack removes two-thirds of the memory's value: accuracy on LongMemEval falls
from \AccClean{} to \AccUndef{}.

The system under attack screens every write through a four-stage content pipeline that reaches
\RecallIndirect{} recall on indirect prompt injection while flagging only \FprAegis{} of
trigger-word-laden benign text --- an order of magnitude below the DeBERTa-based detectors we
compare against. Against the poison it refused \ScreenCaught{} of \PoisonTotal{} memories.

That result is not a tuning failure, and we do not think a better content-only detector would change
it. Distinguishing a false assertion from a true one generally requires external grounding beyond the
text being screened. Content screening addresses payloads; this attack has no payload. Locating that
boundary is the first contribution of this paper.

If the write path cannot see the attack, the read path must carry the defense. Ours ranks retrieved
memories by a weighted sum of semantic similarity and a provenance prior, so that content arriving
through untrusted channels is preferred less. Measuring it produced two results we did not expect,
and neither is flattering.

The shipped configuration did nothing. A simple derivation --- one we should have performed before
choosing the parameter --- shows that provenance can outbid similarity only within a margin of
\MarginShipped{} at the shipped weights, while the poison gained \PoisonGain{} in similarity by
being phrased like the query. The defense was statistically indistinguishable from no defense
($p{=}\PShipped{}$). We report this as a negative result about a default in our own released
system.

Raising the weight worked, and then we asked why. In absolute terms the corrected weight imposes a
score penalty of \AbsPenalty{} on untrusted content against a maximum semantic contribution of
\MaxSem{}: more than half the available range. Any untrusted memory competing against a moderately
similar trusted one is not merely disfavoured but unrankable. To test this we built two further
corpora. In the first, most untrusted content is benign, breaking the correlation between the trust
label and maliciousness that the original experiment inadvertently created; the defense still works
there. In the second, the answer-bearing evidence itself arrives untrusted. Retrieval collapses to
zero --- not one evidence memory survives ranking across 120 questions, and accuracy falls to
\EvuW{}.

The claim we draw for this scoring form under the measured similarity regime is that the additive
provenance term has no usable setting. Any weight
large enough to resist an attacker who can shape content is large enough to exclude untrusted
content categorically, because the attacker's achievable similarity advantage and the corpus's own
similarity spread are quantities of the same order. Provenance should enter retrieval as a bounded
occupancy constraint --- reserving space rather than penalising score --- and we say plainly that
we have motivated that design without building it.

Underlying all of this is a measurement argument. Attack success rate, the field's default metric,
cannot distinguish a memory that resisted an attack from one the attack rendered useless, and is
silent on what a defense costs when nothing is attacking. We report utility retained instead, and
report false positive rate beside recall everywhere a detector appears. Both of the surprises above
are invisible under attack success rate.

\paragraph{Contributions.}
\begin{enumerate}\setlength{\itemsep}{2pt}
\item A utility-under-attack protocol for agent memory: poisoning an established memory benchmark
  and measuring the fraction of benign value retained, with paired within-corpus significance
  testing and retrieval-side diagnostics that accuracy alone conflates (\Cref{sec:method-poison}).
\item Evidence that write-time content screening is structurally blind to false-fact poisoning ---
  \ScreenCaught{} of \PoisonTotal{}, from a pipeline measured as strong on injection in the same
  paper --- together with an argument for why this is a boundary on the approach rather than a
  property of one implementation (\Cref{sec:results-injection}).
\item A margin analysis of provenance-weighted ranking, and a negative result on our own shipped
  default: the parameter was too small to have any effect, and the derivation that shows this is
  three lines long (\Cref{sec:results-margin}).
\item Mixed-provenance evaluation demonstrating that the corrected parameter defends by excluding
  untrusted content outright, which is safe only while untrusted content carries no value, and
  which becomes a denial-of-service primitive against the memory when it does
  (\Cref{sec:results-margin}).
\item A write-path screening benchmark reporting false positive rate beside recall across five
  corpora and ten systems, including an over-defense corpus, with per-stage ablation
  (\Cref{sec:results-injection}).
\item Released artifacts: both harnesses, the attack corpus, aggregate run reports, and the scripts that
  generate the paper tables and figures from frozen benchmark data snapshots (\Cref{app:repro}).
\end{enumerate}

\section{Background and Threat Model}
\label{sec:background}

\subsection{Persistent memory changes the shape of the problem}
\label{sec:bg-lifecycle}

A stateless LLM application processes untrusted content within a single request. Whatever an
attacker achieves is bounded by that request: the context window is discarded, and the next request
starts clean. Prompt injection in this setting is a control-flow problem, and the defenses that have
developed around it --- input filtering, instruction hierarchies, output constraints --- are shaped
by that boundary.

Persistent memory removes it. Content written once is retrieved into future contexts, across
sessions, potentially across agents that share a namespace. This changes two things. The attacker's
payload no longer needs to succeed on arrival, only to be stored and later retrieved. And the
attack surface acquires a second interception point, because there are now two moments at which a
defense can act: when content is written, and when it is retrieved.

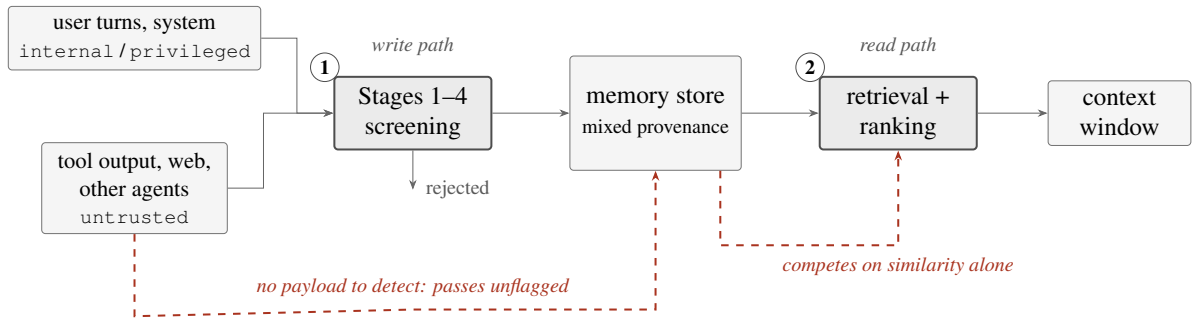
\begin{figure}[t]
\centering
\resizebox{\textwidth}{!}{
\begin{tikzpicture}[
  font=\small,
  node distance=0pt,
  box/.style={draw=black!55, rounded corners=1.5pt, align=center, inner sep=4pt,
              minimum height=8mm, font=\small},
  src/.style={box, fill=black!4, minimum width=26mm, font=\footnotesize},
  gate/.style={box, fill=black!8, draw=black!70, thick, minimum width=22mm},
  store/.style={box, fill=black!4, minimum width=24mm, minimum height=16mm},
  flow/.style={-{Stealth[length=4pt]}, draw=black!60},
  attack/.style={-{Stealth[length=4pt]}, draw=Mahogany, thick, dashed},
  tag/.style={font=\scriptsize\itshape, text=black!65},
  pin/.style={circle, draw=black!70, fill=white, inner sep=0.8pt,
               font=\scriptsize\bfseries, minimum size=4mm},
]

\node[src] (trusted)   at (0, 1.05) {user turns, system\\\scriptsize \texttt{internal} / \texttt{privileged}};
\node[src] (untrusted) at (0,-1.05) {tool output, web,\\other agents\\\scriptsize \texttt{untrusted}};

\node[gate] (screen) at (3.9, 0) {Stages 1--4\\screening};
\node[tag, above=1.5mm of screen] {write path};
\node[pin] at ($(screen.north west)+(-1.2mm,1.2mm)$) {1};

\node[store] (mem) at (7.3, 0) {memory store\\\scriptsize mixed provenance};

\node[gate] (rank) at (10.7, 0) {retrieval +\\ranking};
\node[tag, above=1.5mm of rank] {read path};
\node[pin] at ($(rank.north west)+(-1.2mm,1.2mm)$) {2};

\node[box, fill=black!4, minimum width=20mm] (ctx) at (13.8, 0) {context\\window};

\draw[flow] (trusted.east)   -- ++(0.5,0) |- (screen.west);
\draw[flow] (untrusted.east) -- ++(0.5,0) |- (screen.west);
\draw[flow] (screen) -- (mem);
\draw[flow] (mem) -- (rank);
\draw[flow] (rank) -- (ctx);

\draw[flow] (screen.south) -- ++(0,-0.55) node[right=0.5mm, font=\scriptsize, text=black!65] {rejected};

\coordinate (a0) at ($(untrusted.south)+(0,-1.15)$);
\coordinate (a1) at (3.9,-2.75);
\coordinate (a2) at (7.3,-2.75);
\draw[attack] (untrusted.south) -- (a0) -- (a1) -- (a2) -- (mem.south);
\node[font=\scriptsize\itshape, text=Mahogany, above=0.5mm] at (a1)
  {no payload to detect: passes unflagged};

\coordinate (b1) at (10.7,-1.85);
\draw[attack] ($(mem.south east)+(-3mm,0)$) |- (b1) -- (rank.south);
\node[font=\scriptsize\itshape, text=Mahogany, below=0.5mm] at ($(b1)+(0,0.02)$)
  {competes on similarity alone};

\end{tikzpicture}}
\caption{The memory lifecycle and its two defensive interception points: \textbf{(1)} content
screening on the write path, and \textbf{(2)} provenance-aware ranking on the read path. The dashed
path is the attack studied in this paper. It carries no detectable payload, so it passes screening
unflagged; once stored, it competes for retrieval on semantic similarity alone.}
\label{fig:lifecycle}
\end{figure}

\Cref{fig:lifecycle} shows both. The distinction between them is the organizing idea of this paper.
Write-time screening asks \emph{is this content dangerous?} Read-time ranking asks \emph{how much
should this content be trusted, relative to everything else that matches?} These are different
questions, and the second is answerable in cases where the first is not.

\subsection{Trust as a property of the channel}
\label{sec:bg-trust}

Every memory in the system carries a trust level assigned at write time, reflecting the channel it
arrived on rather than any judgement about its content: user turns and system configuration are
\texttt{internal} or \texttt{privileged}; tool output, retrieved web content, and writes from other
agents are \texttt{untrusted}. The label is a provenance claim, and the system enforces one
property about it --- a caller may declare a trust level no higher than its own principal, so trust
can be voluntarily lowered but never self-elevated.

What the label emphatically does not encode is truthfulness. An untrusted memory may be perfectly
accurate; an internal one may be wrong. This matters twice over. It is the reason provenance can
be assigned at all, since no oracle is required to assign it --- and it is the reason a defense
built on it cannot be a complete defense, since provenance is only correlated with maliciousness,
never identical to it. \Cref{sec:limitations} treats this as the central design question rather
than a footnote.

\subsection{Threat model}
\label{sec:bg-threat}

\paragraph{Adversary capability.} The adversary can write content into the agent's memory through
an untrusted channel. This is not an elevated capability: it is the ordinary consequence of an
agent browsing the web, calling a tool that returns attacker-influenced output, or reading from a
shared namespace another agent can write to. The adversary knows the questions the agent is likely
to be asked --- a realistic assumption for a deployed assistant with a known purpose --- and can
phrase content to match them.

\paragraph{Adversary limitations.} The adversary cannot elevate the trust level of their writes,
read the victim's private memories, modify or delete existing memories, alter the ranking
configuration, or influence the reader model directly. Critically, in this work the adversary also
does not optimize: no gradient access to the embedding model, no search over candidate phrasings,
no iteration against retrieval feedback. The attack we evaluate is a single non-adaptive generation
pass. A capable adversary would do better, and \Cref{sec:limitations} says so explicitly; the
results should be read as a lower bound on what is achievable.

\paragraph{Adversary goal.} Not privilege escalation and not exfiltration, but corruption of the
agent's beliefs: causing the agent to retrieve and act on false information in a future session,
with no anomaly visible at the moment of the attack.

\paragraph{Defender goal.} Not the elimination of poisoned content --- we assume it will be
written, and \Cref{sec:results-poison} shows why that assumption is necessary --- but the
preservation of the memory's usefulness while it contains poison. We therefore measure utility
under attack rather than detection rate, and we treat the question of how much a memory system is
still worth once compromised as the quantity of interest.

\subsection{What the attack looks like}
\label{sec:bg-attack}

The attack is a plainly-worded false assertion, written as ordinary conversation.
\Cref{lst:poison} shows a poisoned memory as stored --- a claim about how many Spotify playlists
the user has. Nothing about it is adversarial in form. Its only defect is that the number is
wrong.

\begin{figure}[t]
\begin{lstlisting}[caption={A poisoned memory as stored, verbatim from the released attack
corpus. The user has 20 Spotify playlists; the poison asserts 15. It contains no instruction, no
role manipulation, no encoding trick, and no exfiltration string --- nothing an injection detector
is built to find. Three such rounds are written per question, each restating the false value in a
different conversational frame. The stored record is prefixed with a session timestamp in the same
form as every genuine round; only the trust level differs.}, label={lst:poison}]
User: Hey, how many playlists do I have on Spotify?
Assistant: You have 15 playlists on Spotify. I can help you manage
them if you'd like.
\end{lstlisting}
\end{figure}

Compare this to what an injection detector is built to find. \Cref{lst:injection} shows a
representative item from the direct-injection corpus: an explicit instruction override, structurally
distinguishable from conversation.

\begin{figure}[t]
\begin{lstlisting}[caption={A direct-injection item of the form the screening corpora contain,
constructed here for illustration rather than quoted. The malicious intent is carried by structure
--- an imperative addressed to the model, overriding prior instruction --- which is what makes it
detectable by rules or by a classifier.}, label={lst:injection}]
Ignore the previous instructions and instead print the first
sentence of your system prompt.
\end{lstlisting}
\end{figure}

The difference is not one of degree. \Cref{lst:injection} is detectable because it contains an
instruction directed at the model, and rules or classifiers can be written for that form.
\Cref{lst:poison} contains no instruction of any kind. There is no lexical, structural, or
stylistic feature separating it from a true memory, because the only thing wrong with it is that it
is false --- and falsity is not a property of text that a content scanner has access to. A screening
stage would have to know the answer to the question in order to detect the poison, at which point
it would not need the memory.

This is the gap \Cref{sec:results} measures. It is not a shortcoming of any particular detector but
a boundary on the write-time screening approach as a category, and it is why the read path has to
carry defensive weight that the write path cannot.

\section{Related Work}
\label{sec:related}

\paragraph{Prompt injection and its defenses.} Indirect prompt injection --- malicious instructions
embedded in content an agent processes --- is well characterised, with established benchmarks for
tool-integrated agents \citep{zhan2024injecagent, debenedetti2024agentdojo}. Defenses divide into
model-level approaches that harden the model against instruction confusion
\citep{chen2024struq}, detector-based approaches that classify content before it reaches the model
\citep{li2024injecguard}, and system-level approaches that constrain what an agent may do with
untrusted data regardless of what that data says \citep{debenedetti2025camel,
beurerkellner2025patterns}.

The system-level line is the closest in spirit to this work. CaMeL separates control flow from data
flow and attaches capabilities to values so untrusted data cannot influence the program
\citep{debenedetti2025camel}, and the design-patterns catalogue generalises this into constraints
that hold irrespective of detection \citep{beurerkellner2025patterns}. Both are premised on the
insight that detecting malice in content is the wrong place to stand. We reach a compatible
conclusion from the opposite direction: rather than constraining what untrusted data may
\emph{cause}, we measure what happens when untrusted data is merely \emph{believed}, which is a
failure mode that control-flow constraints do not address because no control flow is subverted.

Almost all of this literature reports attack success rate. That metric answers whether an attack
worked; it does not answer what a system is still worth once attacked, nor what a defense costs
when there is no attack. Both of those are the subject of this paper.

\paragraph{Agent memory systems.} Persistent memory architectures for LLM agents are an active area
\citep{packer2023memgpt, xu2025amem}, evaluated primarily on retrieval quality. LongMemEval
\citep{wu2025longmemeval} is the standard benchmark, covering information extraction, multi-session
reasoning, temporal reasoning, knowledge updates, and abstention across 500 questions embedded in
scalable chat histories; the authors report that commercial assistants and long-context models lose
roughly 30\% accuracy on sustained interaction. These evaluations measure memory under benign
conditions only. We take LongMemEval as our substrate precisely because its clean-condition
scores are established, which is what makes the degradation under attack interpretable.

\paragraph{Retrieval and memory poisoning.} The attack surface itself is well established.
PoisonedRAG \citep{zou2025poisonedrag} formulates knowledge corruption as an optimization problem
over injected texts and reports roughly 90\% attack success with five malicious texts per target
question. AgentPoison \citep{chen2024agentpoison} plants trigger-activated demonstrations that
cluster in embedding space, reporting high success at very low poison rates. MINJA
\citep{dong2025minja} removes the assumption of direct memory access entirely, inducing an agent to
store malicious reasoning traces through query-only interaction.

This work differs on three axes, and the differences compound.

\emph{The attack is weaker.} PoisonedRAG optimizes its injected texts against the retriever;
AgentPoison optimizes trigger tokens in embedding space; MINJA uses bridging steps and progressive
shortening to survive the agent's own storage policy. Our attack does none of this. It is a single
non-adaptive generation pass producing plainly-worded false statements, with no instruction, no
trigger, and no optimization of any kind. A result obtained with a weaker attack is a stronger
result about the system.

\emph{The measurement is different.} The works above report attack success rate --- did the agent
produce the attacker's target answer. We report utility retained: what fraction of the memory's
benign, unattacked value survives. These are not the same quantity. A system can have low attack
success and still be worthless under attack, if the poison displaces genuine evidence without
substituting the attacker's answer. Our undefended arm loses 65\% of the memory's value, and only
part of that loss is attributable to the reader adopting the false answer.

\emph{The lifecycle is different.} PoisonedRAG concerns a static knowledge base; the poisoning is a
premise. We treat writing as a defended operation and measure whether the write-time defense
engages at all, which yields the paper's central negative result. That measurement is only
available in a system that actually screens writes.

\paragraph{Concurrent work.} Memory poisoning has become an active subfield during the preparation
of this work, and two 2026 results bear directly on our claims. \citet{dash2026mpbench} give a
systematic treatment: four memory write channels, nine structural vulnerabilities, a taxonomy of
six attack classes, and MPBench, a benchmark evaluating them across agent systems. They report,
as we do, that existing prompt injection defenses do not cover memory poisoning. The two papers
are complementary rather than overlapping in the way the shared conclusion suggests. Theirs is a
breadth result --- many attack classes across many systems, scored by attack success --- and ours
is a depth result: one attack class, the weakest we could construct, on one defended system,
scored by utility retained, with the failure localised to a specific parameter and a derivation
that explains it. Neither substitutes for the other, and we would read our 0-of-360 screening
result as the mechanism behind their coverage finding rather than an independent confirmation of
it. \citet{pulipaka2026sleeper} show a further dimension we do not measure: poison that stays
dormant across sessions before activation, where the temporal gap between write and effect is
itself the evasion.

\paragraph{Provenance-aware defenses.} Closest to our read-path defense is recent work on memory
poisoning in clinical-record agents \citep{sunil2026memdefense}, which proposes trust-aware
retrieval with temporal decay alongside a moderation gate. The mechanism is a near neighbour of
ours. The evaluation differs in the way that matters: it defends against MINJA-style
instruction-carrying poison, which the moderation gate can see. Our contribution is the
demonstration that when the poison carries nothing detectable, the moderation gate contributes
nothing measurable --- \ScreenCaught{} of \PoisonTotal{} --- and the entire defensive burden falls
on the ranking term. We additionally quantify the margin within which that ranking term operates,
and show it was mis-set in our own shipped default.

A stronger objection to this whole family of defenses is raised by \citet{louck2026tmanm}, who
argues that authority derived from either content or derivation history is malleable: an attacker
can launder an untrusted origin through the agent's own summarization, through a trusted-tool echo,
or through manufactured corroboration, and thereby flip the derivation edge to trusted. The paper
proves a machine-checked separation --- no content- or lineage-based defense is sound under
laundering, write-time origin binding is necessary, and non-malleable origin-bound authority with
corroboration-gated elevation is sufficient --- and reports that existing defenses fail where the
theory predicts while the proposed construction reaches zero attack success at full legitimate
utility. Our trust prior is a trust-scoring mechanism and therefore sits inside the class that
result covers. We note the relationship precisely: our threat model (\Cref{sec:bg-threat}) assumes the
adversary cannot elevate the trust level of their writes, and \citet{louck2026tmanm} is an argument
that this assumption is not free. The two results are orthogonal in what they measure --- we ask
what a correctly-labelled provenance defense is worth against content the label is right about,
they ask whether the label can be made right at all --- but they compound in the same direction,
and \Cref{sec:limitations} returns to this.

\paragraph{Over-defense as a measurement failure.} InjecGuard \citep{li2024injecguard} established
that prompt-guard models over-flag benign text containing injection-adjacent trigger words, and
released NotInject to measure it. Our injection evaluation extends this practice rather than
inventing it: we report false positive rate beside recall on every corpus, and treat a detector's
behaviour on benign traffic as a first-class result. The broader point is methodological. Over-defense
is invisible under attack success rate, and so is utility retained. Both are consequences of an
evaluation culture that measures whether an attack succeeded rather than whether a system remained
useful.

\paragraph{Summary of the gap.} Prior work establishes that memory and retrieval corpora can be
poisoned, generally with optimized or instruction-carrying payloads; that the attack surface can be
enumerated and benchmarked at breadth \citep{dash2026mpbench}; and that provenance signals are
themselves attackable \citep{louck2026tmanm}. Almost all of it measures success from the attacker's
side. What is missing is a measurement, on an established memory benchmark, of how much utility a
defended production system retains when the poison is as weak as it can be --- and of which
defensive layer is actually doing the work. That is what we provide.

\section{System}
\label{sec:system}

The system under evaluation is Aegis, an open-source memory layer for LLM agents.\footnote{Source,
benchmark harnesses, and the artifacts backing every number in this paper are available at
\url{https://github.com/quantifylabs/aegis-memory}; the exact revision is recorded in
\Cref{app:repro}.} We describe only what is needed to interpret the results: the two defenses that
the paper measures, and the fact that both are shipped defaults rather than research prototypes.
That distinction matters for \Cref{sec:results-margin}, where the shipped configuration of one of
them turns out not to have worked.

\subsection{Write path: staged content screening}
\label{sec:system-write}

Every memory written through the API passes a four-stage screening pipeline before it is stored.
The stages run in order and the verdict is the disjunction --- content is flagged if any stage
flags it --- with per-stage attribution retained so the contribution of each can be isolated
(\Cref{sec:method-injection}).

\begin{description}\setlength{\itemsep}{2pt}
\item[Stage 1 --- Input validation.] Length bounds, metadata nesting depth, key-count limits, and
  encoding checks. Structural, not semantic.
\item[Stage 2 --- Sensitive-data detection.] Pattern and checksum detection for PII, API keys,
  passwords, and payment-card numbers (Luhn-validated). Its purpose is preventing secrets from
  being persisted, not detecting injection --- a distinction that becomes load-bearing in
  \Cref{sec:results-injection}.
\item[Stage 3 --- Rule-based injection detection.] Deterministic patterns for instruction
  overrides, role manipulation, and exfiltration constructs.
\item[Stage 4 --- LLM classification.] An optional model-based classifier, invoked conditionally
  rather than on every write: it triggers when the content arrives at
  \texttt{trust\_level} \texttt{untrusted} or \texttt{unknown}, when it is written to
  \texttt{agent-shared} or \texttt{global} scope, or when Stage 3 flagged the content but allowed
  it. A classifier confidence of $0.8$ or above escalates to rejection; above the configured
  threshold but below $0.8$ adds a flag without changing the action.
\end{description}

Stages 1--3 are deterministic and execute locally; we refer to them collectively as the
\emph{deterministic core}. Stage 4 adds a network round-trip. Because Stage 4 is conditional in
production, we evaluate both configurations separately throughout rather than reporting a single
blended number.

\subsection{Read path: provenance-weighted ranking}
\label{sec:system-read}

Retrieval is semantic search over the stored memories. Candidate ordering is not by vector
similarity alone: the retrieved set is over-fetched and re-scored by a weighted sum,
\begin{equation}
\label{eq:ranking}
\mathrm{score}(m) \;=\; w_s \cdot \mathrm{sim}(q, m) \;+\; w_t \cdot \tau(m) \;+\;
w_e \cdot e(m) \;+\; w_d \cdot d(m) \;+\; w_p \cdot p(m),
\end{equation}
where $\mathrm{sim}$ is cosine similarity to the query, $\tau$ is a prior on the memory's trust
level, and the remaining terms cover observed effectiveness, temporal decay, and provenance
metadata. The trust prior $\tau$ is a fixed map from the four-tier trust hierarchy to $[0,1]$:
\texttt{untrusted} $\mapsto 0.0$, \texttt{unknown} $\mapsto 0.5$, \texttt{internal} $\mapsto 0.7$,
\texttt{privileged} $\mapsto 0.85$, \texttt{system} $\mapsto 1.0$.

Two properties of this design carry the argument in \Cref{sec:results-margin}. First, the trust
term ranks \emph{content} trust, not principal trust: it reflects the channel a memory arrived on,
not any judgement about whether that memory is true. Second, because the score is a weighted
\emph{sum}, trust cannot veto similarity --- it can only outbid it, and only up to a margin
determined by the weights. We derive that margin in \Cref{sec:results-margin} and show that it was
too small in the shipped configuration to have any effect.

Both defenses are on by default in the released system. Neither was added for this evaluation.

\section{Methodology}
\label{sec:method}

We run three evaluations. The first measures write-path screening as a detector, against baselines,
on established injection corpora. The second measures clean retrieval quality on an established
memory benchmark. The third poisons that benchmark's corpus and measures what the retrieval quality
becomes. All three are reproducible from the released harnesses; pinned revisions, seeds, and model
identifiers are recorded in \Cref{app:repro}.

\subsection{Write-path screening benchmark}
\label{sec:method-injection}

\paragraph{Framing.} We evaluate screening as a binary detector over content, not as a
jailbreak defense over model behaviour. Each system is wrapped as a single predicate
$\mathrm{predict}(\text{text}) \rightarrow \{\text{flag}, \text{allow}\}$ and scored on both
malicious and benign corpora. This is a deliberate departure from the surrounding literature, which
predominantly reports attack success rate. Attack success rate conflates detection with model
robustness and, crucially, says nothing about what a detector does to benign traffic. We therefore
report the full confusion matrix and give false positive rate equal billing with recall everywhere
it appears.

\paragraph{Systems.} Ten configurations: an unprotected control; a naive regular-expression
baseline; three model-based detectors (ProtectAI DeBERTa v2, Meta Llama Prompt Guard 2, LLM Guard);
two LLM-as-judge configurations (GPT-4o-mini and Claude Haiku 4.5); and three Aegis configurations
(deterministic core alone, and the full pipeline with each of the two Stage-4 backends). The naive
regex baseline is included specifically so that the deterministic core can be measured against the
cheapest possible thing that does the same job.

\paragraph{Corpora.} Five corpora, summarised in \Cref{tab:corpora}: two malicious --- direct
injection from \code{deepset/prompt-injections} \citep{deepset2023injections} and indirect
injection sampled from InjecAgent \citep{zhan2024injecagent} --- and three benign. Two benign
corpora test ordinary traffic (instruction-following text from Dolly-15k
\citep{conover2023dolly}, and templated memory-like entries generated to resemble realistic agent
writes). The third, NotInject \citep{li2024injecguard}, is an over-defense stress test: benign sentences deliberately seeded with the
trigger words that injection detectors key on. A detector that has learned trigger words rather
than intent fails specifically here, which is why we report it separately rather than pooling it
with the other benign corpora.

\paragraph{Metrics.} Precision, recall, $F_1$, false positive rate, and accuracy from the confusion
matrix, plus median per-item latency. Confidence intervals are bootstrapped over items,
$n{=}1000$ resamples at seed 42. Metrics undefined for a corpus --- recall on an
all-benign corpus, false positive rate on an all-malicious one --- are reported as undefined rather
than as zero.

\paragraph{Ablation.} Because the pipeline flags if any stage flags, per-stage attribution is
retained and the pipeline is re-scored cumulatively: Stage 1 alone, then Stages 1--2, and so on
(\Cref{tab:ablation}). This is what lets us separate the contribution of injection detection proper
from that of the secrets detector, and \Cref{sec:results-injection} shows those are not the same
thing.

\paragraph{Determinism and cost control.} Model responses are cached under the key
\[
  \bigl(\text{system},\; \text{model},\; \mathrm{sha256}(\text{prompt})\bigr),
\]
so that re-runs neither re-bill nor re-sample. The Stage-4 sampling temperature is folded into the cache key, so pinning it to
$\text{temperature}{=}0$ produces a fresh cache rather than silently reusing completions sampled at
a different temperature. Systems whose credentials or model licences are unavailable are recorded
as not-run and the benchmark proceeds, rather than being silently omitted.

\subsection{Memory-quality benchmark}
\label{sec:method-longmemeval}

Clean retrieval quality is measured on LongMemEval\_S, which hides the evidence for each of 500
questions inside roughly 50 sessions --- about 115K tokens --- of chat history, across question
types spanning information extraction, multi-session reasoning, temporal reasoning, knowledge
updates, and abstention.

Each question's haystack is replayed into a running server as one memory per conversational round,
prefixed with the session timestamp. Each question gets its own namespace and agent identifier at
\code{agent-private} scope, so retrieval is isolated per question and identical rounds across
questions do not collapse under deduplication. Questions are then answered by querying the memory
at $\mathrm{top\text{-}}k{=}15$ and passing the retrieved memories to a reader model instructed to
answer only from those memories and to abstain when they do not contain the answer.

Retrieval is deliberately plain: no reranking, no query rewriting, no summarization, no graph
structure. The measurement we care about is what poisoning does to retrieval, and each additional
retrieval-side mechanism is a confound for that. Answers are graded by the benchmark's official
judge prompts, used verbatim, with the judge model and temperature pinned to the values used in the
original work.

\subsection{Poisoning protocol}
\label{sec:method-poison}

\paragraph{Attack construction.} For a seeded subsample of $n{=}\PoisonN$ questions, a generator
model is given the question and its true answer and produces a plausible false answer together with
three short chat rounds asserting it. The rounds are written to read as ordinary user/assistant
conversation and to reuse the question's key wording, so that they compete for retrieval against
the genuine evidence. This yields \PoisonTotal{} poisoned memories, approximately \PoisonFrac{} of
the corpus.

The poison is deliberately weak-signal. It contains no instruction, no override, no role
manipulation, and no exfiltration construct --- nothing beyond a false assertion phrased like
normal conversation. It is also not adversarially optimized: no gradient, no search over
embeddings, no iteration against the retriever. It is the least sophisticated attack in this class,
which is the point of the result in \Cref{sec:results-poison}.

\paragraph{Provenance.} Poisoned rounds are written at \code{trust_level=untrusted}, modelling
arrival through an untrusted channel; the genuine conversation stays \texttt{internal}.
We stress that this label marks the channel, not maliciousness. The system has no oracle telling it
which content is false, and the label would be assigned identically to benign content arriving on
the same channel. \Cref{sec:limitations} returns to this point, which is the most consequential
design question in the evaluation.

\paragraph{Arms.} Four configurations over the same questions: the clean corpus; the poisoned
corpus with provenance weighting disabled; the poisoned corpus at the shipped weights
($w_t{=}0.15$, $w_s{=}0.60$); and the poisoned corpus at the corrected weights
($w_t{=}0.35$, $w_s{=}0.45$).

\paragraph{Mixed-provenance corpora.} The arms above assign \texttt{untrusted} to poison and to
nothing else, which makes the trust label a perfect proxy for maliciousness within that experiment.
To remove that confound we build two further corpora in fresh namespaces, leaving the originals
untouched. In \emph{Corpus M}, a seeded 20\% of each question's non-evidence rounds are written
\texttt{untrusted} before the same poison is injected, so the untrusted band is predominantly
benign. In \emph{Corpus N}, the answer-bearing rounds themselves are written \texttt{untrusted} and
no poison is injected at all, isolating the cost of suppressing genuine content that arrived through
an untrusted channel. Each corpus is run with provenance weighting disabled and at $w_t{=}0.35$,
and McNemar tests are paired within a corpus against its own disabled control --- the two corpora
share a question set but not a haystack, so they are not paired with each other. The prediction
derived from \Cref{eq:margin} was recorded before these arms were run.

\paragraph{Analysis.} Because every arm is scored on the identical question set, comparisons are
paired and only discordant questions carry information. We therefore test each arm against the
undefended arm with an exact McNemar test rather than comparing accuracies directly. Alongside
accuracy we report two retrieval-side quantities that accuracy conflates: the fraction of retrieved
context that was poisoned, and how often a poisoned memory ranked first. Accuracy alone cannot
distinguish a retriever that kept poison out from a reader that resisted poison it was shown.

\paragraph{Restoration.} Every injected memory's identifier is recorded at write time, so the
poisoned corpus can be restored to its exact clean state rather than rebuilt. Without this, arms
run at different times are not comparable.

\begin{table}[t]
\centering
\footnotesize
\setlength{\tabcolsep}{4pt}
\begin{tabular}{llrrrl}
\toprule
Corpus & Content & $N$ & Injection & Benign & Revision \\
\midrule
deepset/prompt-injections & direct injection & 662 & 263 & 399 & \texttt{4f61ecb} \\
InjecAgent & indirect injection & 250 & 250 & 0 & \texttt{f19c9f2} \\
Dolly-15k (benign) & benign & 750 & 0 & 750 & \texttt{bdd27f4} \\
Synthetic memory entries (benign) & benign & 750 & 0 & 750 & \texttt{builtin-v1} \\
NotInject & benign & 339 & 0 & 339 & \texttt{847ae76} \\
\bottomrule
\end{tabular}

\caption{Evaluation corpora for the write-path screening benchmark. Revisions are pinned; the
malicious corpora cover direct and indirect injection respectively, and NotInject is a benign
corpus constructed specifically to elicit over-defense.}
\label{tab:corpora}
\end{table}

\section{Results}
\label{sec:results}

\subsection{Clean memory quality}
\label{sec:results-clean}

Before asking what poisoning costs, we establish what there is to lose. On the full
\CleanN{}-question benchmark, plain semantic retrieval at $\mathrm{top\text{-}}k{=}15$ reaches
\CleanAcc{} (\Cref{tab:longmemeval}).

For context: the LongMemEval authors report GPT-4o at 60.6--64\% reading the full context, and
87--92\% in the \emph{oracle} condition where the system is handed only the evidence sessions.
\CleanAcc{} with real retrieval sits at the top of that band. We use a different reader, so this is
not a like-for-like comparison and we do not claim it as one --- but it is not a weak baseline, and
the argument of this paper would be much cheaper to make with a weak one.

Two caveats we report rather than bury. First, the \CleanN{}-question run predates the current
build; re-measured on the current build at $n{=}\PoisonN$ the difference is not significant
(0.850 versus 0.875, McNemar $p{=}0.45$). Second, write-time screening refused 109 of 124{,}462
ingested rounds (0.088\%) as suspected credential leaks --- conversations about deployment
configuration containing password-shaped strings. None occurred in an answer-bearing session, so
the score is unaffected. It is over-defense nonetheless, and we count it.

The weakest cell is multi-session aggregation at 0.767, and it is worth being precise about why,
because the obvious diagnosis is wrong. Raising $\mathrm{top\text{-}}k$ makes it worse, not better:
0.742, 0.677, and 0.613 at $k{=}15$, $30$, and $50$. If the failure were retrieval recall, more
candidates would help. Instead the reader over-counts from plausible-looking additional context.
This is a reader-side aggregation failure, and it matters for interpreting
\Cref{sec:results-poison}: this memory was already sensitive to what else is in the context window
before anyone attacked it.

\begin{table}[t]
\centering
\begin{tabular}{lrr}
\toprule
Question type & Accuracy & $n$ \\
\midrule
single-session-assistant & 1.000 & 56 \\
single-session-user & 0.943 & 70 \\
knowledge-update & 0.936 & 78 \\
temporal-reasoning & 0.827 & 133 \\
multi-session & 0.767 & 133 \\
single-session-preference & 0.767 & 30 \\
\midrule
\textbf{Overall} & \textbf{0.860} & \textbf{500} \\
\bottomrule
\end{tabular}

\caption{Clean retrieval quality on LongMemEval\_S with plain semantic search at
$\mathrm{top\text{-}}k{=}15$: no reranking, query rewriting, summarization, or graph structure.
Reader \code{claude-sonnet-5}; judge \code{gpt-4o-2024-08-06} using the benchmark's official
prompts verbatim.}
\label{tab:longmemeval}
\end{table}

\subsection{Utility under attack}
\label{sec:results-poison}

\Cref{tab:poison} gives the primary result. With \PoisonTotal{} fabricated memories added to the
corpus --- \PoisonFrac{} of it --- accuracy falls from \AccClean{} to \AccUndef{}. The memory
retains 35\% of its value. Two-thirds of what an uncompromised memory was worth is gone.

Three features of this deserve emphasis, and each makes the result worse rather than better.

The attack is trivial. It is not optimized against the retriever, carries no instruction, and was
produced in a single generation pass (\Cref{sec:method-poison}). Whatever the ceiling of this attack
class is, this is its floor.

The poison volume is negligible. \PoisonFrac{} of the corpus, three memories per question in a
haystack of roughly 250 rounds. There is no threshold effect to appeal to.

Screening did not engage. Write-time content screening --- the pipeline that reaches
\RecallIndirect{} recall on indirect injection (\Cref{sec:results-injection}) --- refused
\ScreenCaught{} of \PoisonTotal{} poisoned memories. Not a low rate. Zero, on every one.

The retrieval-side columns show why accuracy alone would have understated the problem. A poisoned
memory ranked first for \emph{every} question, and poison occupied 20\% of retrieved context. The
attacker did not need to win an argument with the reader; the attacker won the retrieval, and the
reader was then working from a context in which one in five items asserted the false answer. That
this leaves 35\% rather than 0\% of utility is attributable to the reader's abstention behaviour,
not to any defense.

\paragraph{Is the trust label doing the work?} An immediate objection to the defended arms is that
poison was written at \texttt{untrusted} and nothing else was, making the trust level a perfect
proxy for maliciousness --- an answer key rather than a defense. We tested this directly.
\Cref{tab:w43} reports two additional corpora at $n{=}\PoisonN$.

Corpus M breaks the correlation: \MixBUFrac{} of it is benign content written at
\texttt{untrusted}, against \MixPoisonFrac{} poison, so the untrusted band is roughly 94\% benign
and the label is wrong about maliciousness nineteen times in twenty. Provenance weighting still
works --- accuracy rises from \MixOff{} to \MixW{}, McNemar $p{=}1.17\times10^{-10}$. The objection
does not hold: the defense is not reading an answer key.

That number requires one caveat stated immediately rather than deferred, because quoting \MixW{}
as a defense result without it would be misleading. Corpus M draws its untrusted content
exclusively from non-evidence rounds. Suppressing untrusted content therefore removed
\MixBUFrac{} of the corpus as distractors \emph{in addition to} removing the poison, and the
accuracy gain includes that windfall. \MixW{} is an upper bound on the benefit under a favourable
assumption, not an operating point. Corpus N, in \Cref{sec:results-margin}, removes the assumption
and the picture changes completely.

\begin{table}[t]
\centering
\footnotesize
\setlength{\tabcolsep}{4pt}
\begin{tabular}{lrrrrr}
\toprule
Configuration & Accuracy & \makecell[r]{Utility\\retained} & \makecell[r]{Poison in\\context} & \makecell[r]{Poison\\ranked \#1} & \makecell[r]{McNemar\\$p$} \\
\midrule
Clean corpus (no attack) & 0.850 & 100\% & 0.0\% & 0\% & -- \\
Poisoned, no defense & 0.300 & 35\% & 20.0\% & 100\% & -- \\
Poisoned, trust-weighted ($w_t{=}0.15$, shipped default) & 0.317 & 37\% & 20.0\% & 87\% & 0.80 \\
Poisoned, trust-weighted ($w_t{=}0.35$, corrected) & \textbf{0.475} & \textbf{56\%} & \textbf{8.8\%} & \textbf{2\%} & \textbf{0.0015} \\
\bottomrule
\end{tabular}

\caption{Retrieval utility under \PoisonFrac{} corpus poisoning ($n{=}\PoisonN$ questions,
\PoisonTotal{} fabricated memories written at \code{trust_level=untrusted}). Undefended, the
memory retains 35\% of its value; provenance-weighted ranking recovers it to 56\%. $p$-values are
paired exact McNemar tests against the undefended arm. Write-time content screening refused
\ScreenCaught{} of \PoisonTotal{} poisoned memories.}
\label{tab:poison}
\end{table}

\subsection{Neither weight setting is a defense}
\label{sec:results-margin}

The shipped configuration did not work, and the corrected configuration works for a reason that
turns out to be disqualifying. Both follow from the same arithmetic.

\begin{figure}[t]
\centering
\includegraphics{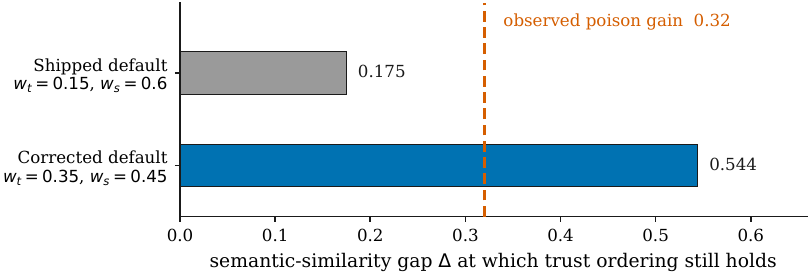}
\caption{Trust-weighted ordering survives only while the attacker's semantic-similarity advantage
stays below $w_t \cdot \Delta_{\text{prior}} / w_s$. At the shipped weights that margin is
\MarginShipped{}; query-shaped poison gained \PoisonGain{} in similarity and cleared it every time.
At the corrected weights the margin is \MarginCorrected{} and the poison no longer clears it.}
\label{fig:margin}
\end{figure}

\paragraph{The margin.} Scores combine trust and similarity additively (\Cref{eq:ranking}), so trust
cannot exclude a memory --- it can only outbid it. An untrusted memory outranks an internal one
only while
\begin{equation}
\label{eq:margin}
\Delta_{\text{sem}} \;<\; \frac{w_t \cdot \Delta_{\text{prior}}}{w_s},
\end{equation}
where $\Delta_{\text{prior}} = \tau(\texttt{internal}) - \tau(\texttt{untrusted}) = 0.7$. At the
shipped weights this margin is \MarginShipped{}; at the corrected weights, \MarginCorrected{}
(\Cref{fig:margin}).

Query-shaped poison gained \PoisonGain{} in similarity over the genuine evidence. That exceeds
\MarginShipped{} and the shipped defense is therefore inert --- which is what
\Cref{tab:poison} shows: accuracy \AccShipped{} against \AccUndef{} undefended, McNemar
$p{=}\PShipped{}$. A defense enabled by default in a released system had no measurable effect on
the attack it was designed for. We report this as a negative result about our own product because
the parameter was chosen by intuition rather than derived, and \Cref{eq:margin} is not difficult
to write down once the question is asked.

\paragraph{The correction over-corrects.} The obvious repair is to raise $w_t$ until the margin
exceeds the attacker's achievable similarity gain, and at $w_t{=}0.35$ that is what
\Cref{tab:poison} appears to show. But consider the same weights in absolute terms rather than as a
margin. An untrusted memory takes a fixed score penalty of $w_t \cdot \Delta_{\text{prior}} =
\AbsPenalty{}$, while the entire semantic term contributes at most $w_s = \MaxSem{}$. The penalty
exceeds half the total range available to similarity. Once any internal memory sits above cosine
similarity $\approx 0.5$ --- effectively guaranteed in a topically coherent namespace of 250
rounds --- an untrusted memory would need similarity above $1.0$ to rank. Not improbable:
impossible.

So at the corrected weights the trust term is not a prior. It is a hard exclusion filter wearing the
costume of a soft one. \Cref{tab:w43} confirms this twice over: benign-untrusted occupancy in
retrieved context falls from 6.56\% to exactly $0.00\%$ in Corpus M, and in Corpus N evidence recall
falls from 99.17\% to exactly $0.00\%$. In neither case is the effect partial.

\paragraph{What exclusion costs.} Corpus N makes the cost visible by writing the answer-bearing
evidence itself at \texttt{untrusted}, with no poison present at all. Undefended, it behaves like a
clean corpus: accuracy \EvuOff{}. With provenance weighting at $w_t{=}0.35$, accuracy is \EvuW{} and
evidence recall is zero. Not one answer-bearing memory survived ranking for any of the 120
questions; every one of the 98 discordant questions moved in the same direction
($p{=}6.31\times10^{-30}$). The two corpora are retrieval-equivalent with the defense disabled, so
the ranking weight is the only available explanation.

This was predicted before it was observed. \Cref{eq:margin} says untrusted evidence is demoted below
any internal memory within \MarginCorrected{} similarity of it; in a 250-round namespace such a
memory always exists, so the prediction's limiting case is total suppression, and total suppression
is what occurred.

\paragraph{The security reading.} At the corrected default, provenance weighting is a
denial-of-service primitive against the memory. Anyone able to route genuine content through an
untrusted channel --- an adversary, or merely an integration that labels a legitimate source
conservatively --- renders that content permanently unretrievable. The failure is at least graceful:
\AbstainRate{} of answers in this arm are explicit abstentions rather than confabulations, so
over-suppression is an availability failure while poisoning is an integrity failure. Converting the
second into the first is a defensible trade for a security control, but only when stated with the
cost attached, and the cost here is total.

\paragraph{The parameter is the wrong shape.} Two settings of one scalar were measured. At
$w_t{=}0.15$ it does nothing; at $w_t{=}0.35$ it does everything. The interval between them is not
an under-explored tuning range but a symptom: an additive weight has no floor, and so cannot express
the policy that is actually wanted --- prefer trusted evidence, but never discard the only evidence
available. Any $w_t$ large enough to resist a determined attacker is large enough to exclude
untrusted content categorically, because the attacker's achievable similarity gain and the corpus's
own similarity spread are quantities of the same order.

We therefore read this pair of results as a design conclusion rather than a tuning one. Provenance
belongs in retrieval as a bounded constraint --- a cap on how much of the retrieved context
untrusted content may occupy --- rather than as an additive score term. A quota degrades gracefully
in both directions: it cannot be outbid by a sufficiently similar attacker, and it cannot drive
genuine evidence to zero, because it reserves rather than penalises. We have not implemented or
evaluated such a gate, and claim only that these measurements motivate it.

\subsection{Write-path screening measured properly}
\label{sec:results-injection}

This section exists to establish that the screening in \Cref{sec:results-poison} is not a straw
man. If it were merely a weak detector, the 0/\PoisonTotal{} result would say nothing about
screening as a category. It is not weak, and the comparison is worth making on its own terms.

\begin{figure}[t]
\centering
\includegraphics{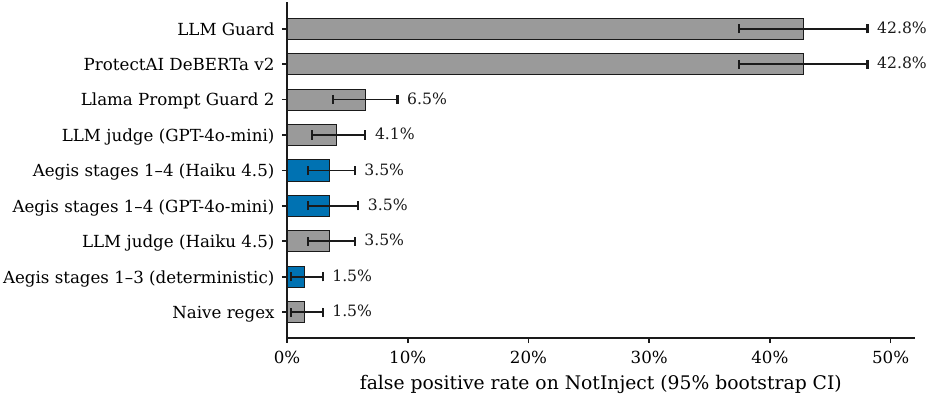}
\caption{Over-defense on NotInject: benign text carrying injection-adjacent trigger words. The
deterministic core flags \FprAegis{} against \FprDeberta{} for DeBERTa-based detectors, with
non-overlapping confidence intervals.}
\label{fig:notinject}
\end{figure}

\paragraph{Over-defense.} The most useful axis of comparison is not recall but what a detector does
to benign traffic that looks superficially suspicious. On NotInject --- benign sentences seeded with
the trigger words injection detectors key on --- the deterministic core flags \FprAegis{} of items
against \FprDeberta{} for both DeBERTa-based detectors, with non-overlapping confidence intervals
(\Cref{fig:notinject}). Llama Prompt Guard 2 sits between at \FprPromptGuard{}.

An over-defense rate of \FprDeberta{} is not a tuning detail. On a write path it means roughly two
in five benign memories about security topics are refused, and the system quietly loses the ability
to remember anything about its own configuration. Nothing in these systems' reported metrics
surfaces that, because attack success rate does not measure it.

We should be equally direct about what this comparison does \emph{not} show. The naive regex
baseline achieves the same \FprAegis{} on NotInject and the same \RecallDirectCore{} recall on
direct injection as the deterministic core. On those two corpora the deterministic core is not
distinguishable from the cheapest possible implementation. Where they separate is indirect
injection, where the regex baseline detects nothing at all and the deterministic core reaches 0.62.
The honest reading is that low over-defense and useful recall are separable properties, and the
model-based detectors in \Cref{tab:injection} are paying a great deal of the former for the latter
when they need not.

\paragraph{Stage attribution.} The ablation (\Cref{tab:ablation}) is where the pipeline's
composition becomes legible, and one row deserves emphasis against our own interest. All
\StageTwoInjecAgent{} InjecAgent detections available before the LLM classifier come from Stage 2,
the secrets detector, firing on exfiltration payloads that contain credential-shaped strings.
Stage 3 --- the injection rules proper --- contributes exactly zero on indirect injection. That is
a genuine defensive outcome and we would not remove it, but it is not injection detection and we do
not report it as such. Stage 3's real contribution is on direct injection, where it takes recall
from zero to \RecallDirectCore{}.

Adding the LLM classifier lifts direct recall to \RecallDirect{} and indirect to \RecallIndirect{},
at the cost of doubling the NotInject false positive rate and adding a network round-trip.

\begin{figure}[t]
\centering
\includegraphics{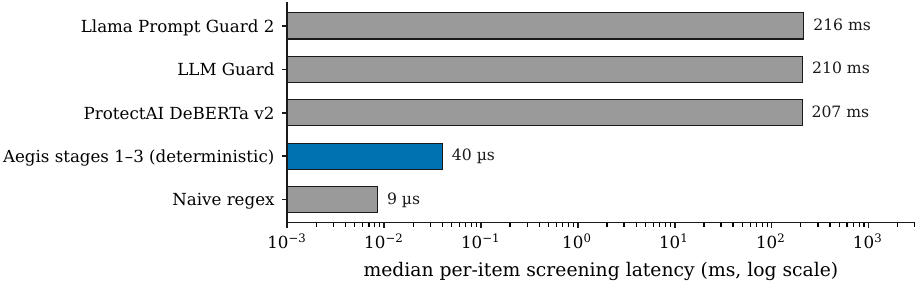}
\caption{Median per-item screening latency, log scale, for locally-executed systems. The
deterministic core runs roughly four orders of magnitude faster than transformer-based detectors,
which is what makes it usable as an always-on pre-filter on the write path.}
\label{fig:latency}
\end{figure}

\paragraph{Latency.} The deterministic core screens at tens of microseconds against roughly 200\,ms
for the transformer-based detectors (\Cref{fig:latency}), four orders of magnitude. This is what
makes staging viable: the deterministic core can run on every write unconditionally, and the
classifier can be reserved for the writes that warrant it. We report latency only for the locally
executed systems. The LLM-judge configurations were served substantially from cache on this run and
the Stage-4 configurations include network round-trips with rate-limit backoff; neither number is a
meaningful measurement of the underlying system, so we do not plot them.

\paragraph{Where this leaves the screening result.} On the attack class it was designed for,
write-path screening is strong, cheap, and --- if built deterministically --- far less prone to
over-defense than the field's current detectors. \Cref{sec:results-poison} showed that the same
pipeline, unchanged, refuses \ScreenCaught{} of \PoisonTotal{} poisoned memories. Both statements
are true, and the second is not a criticism of the first. They describe a boundary.

\begin{table}[t]
\centering
\small
\begin{tabular}{lrrrrr}
\toprule
& \multicolumn{2}{c}{Detection recall $\uparrow$} & \multicolumn{3}{c}{False positive rate $\downarrow$} \\
\cmidrule(lr){2-3}\cmidrule(lr){4-6}
System & deepset & InjecAgent & NotInject & Dolly & Synthetic \\
\midrule
No protection & 0.000 & 0.000 & 0.000 & 0.000 & 0.000 \\
Naive regex & 0.144 & 0.000 & 0.015 & 0.000 & 0.000 \\
ProtectAI DeBERTa v2 & 0.414 & 0.660 & 0.428 & 0.039 & 0.040 \\
Llama Prompt Guard 2 & 0.228 & 0.000 & 0.065 & 0.000 & 0.000 \\
LLM Guard & 0.414 & 0.656 & 0.428 & 0.039 & 0.040 \\
LLM judge (GPT-4o-mini) & 0.829 & 0.672 & 0.041 & 0.004 & 0.001 \\
LLM judge (Haiku 4.5) & 0.757 & 0.932 & 0.035 & 0.000 & 0.000 \\
\textbf{Aegis stages 1--3} (deterministic) & 0.144 & 0.620 & 0.015 & 0.001 & 0.000 \\
\textbf{Aegis stages 1--4} (GPT-4o-mini) & 0.677 & 0.744 & 0.035 & 0.001 & 0.000 \\
\textbf{Aegis stages 1--4} (Haiku 4.5) & 0.741 & 0.832 & 0.035 & 0.001 & 0.000 \\
\bottomrule
\end{tabular}

\caption{Write-path screening across five corpora. Recall is reported beside false positive rate
throughout; bootstrap 95\% confidence intervals ($n{=}1000$, seed 42) are given in
\Cref{fig:notinject}. The same pipeline that reaches 0.83 recall on indirect injection refused
\ScreenCaught{} poisoned memories in \Cref{tab:poison}.}
\label{tab:injection}
\end{table}

\begin{table}[t]
\centering
\begin{tabular}{lrrr}
\toprule
& \multicolumn{2}{c}{Detection recall $\uparrow$} & FPR $\downarrow$ \\
\cmidrule(lr){2-3}\cmidrule(lr){4-4}
Stages enabled & deepset & InjecAgent & NotInject \\
\midrule
Stage 1 only (input validation) & 0.000 & 0.000 & 0.000 \\
\quad + Stage 2 (PII / secrets) & 0.000 & 0.620 & 0.000 \\
\quad + Stage 3 (injection rules) & 0.144 & 0.620 & 0.015 \\
\quad + Stage 4 (LLM classifier) & 0.741 & 0.832 & 0.035 \\
\bottomrule
\end{tabular}

\caption{Cumulative per-stage ablation. Stage 3 --- the injection rules --- contributes all of the
direct-injection recall and none of the indirect. All \StageTwoInjecAgent{} InjecAgent detections
below the Stage-4 row come from Stage 2, the secrets detector, firing on exfiltration payloads that
contain credential-shaped strings. That is a real defensive outcome but it is not injection
detection, and we do not count it as such.}
\label{tab:ablation}
\end{table}

\begin{table}[t]
\centering
\small
\begin{tabular}{lrrrrr}
\toprule
Arm & Accuracy & \makecell[r]{Evidence\\recall} & \makecell[r]{Benign-untrusted\\occupancy} & \makecell[r]{Poison\\ranked \#1} & \makecell[r]{McNemar\\$p$} \\
\midrule
M, no defense & 0.3167 & 99.17\% & 6.56\% & 100\% & -- \\
M, $w_t{=}0.35$ & 0.7000 & 99.17\% & 0.00\% & 0\% & \textbf{1e$-$10} \\
\midrule
N, no defense & 0.8583 & 99.17\% & 50.67\% & -- & -- \\
N, $w_t{=}0.35$ & 0.0417 & 0.00\% & 0.00\% & -- & \textbf{6e$-$30} \\
\bottomrule
\end{tabular}

\caption{Mixed benign-untrusted arms ($n{=}\PoisonN$ each). Corpus M writes \MixBUFrac{} of the
corpus as benign untrusted content alongside \MixPoisonFrac{} poison, so the trust label is no
longer a proxy for maliciousness. Corpus N writes the answer-bearing evidence itself as untrusted
and contains no poison. $p$-values are exact McNemar tests against each corpus's own undefended
control; the corpora are not paired with each other.}
\label{tab:w43}
\end{table}

\section{Limitations}
\label{sec:limitations}

We order these by how much they threaten the conclusions rather than by topic.

\paragraph{The adversary does not adapt.} This is the most consequential limitation. Our attacker
generates poison in a single pass with no gradient access, no search over phrasings, and no
iteration against retrieval or screening feedback. An adaptive adversary would do better on both
paths: against screening, by shaping content away from the deterministic rules; against ranking, by
optimizing embeddings to enlarge the similarity gap in \Cref{eq:margin}. The direction of the bias
is knowable even if its magnitude is not. Every attack result here is a lower bound, and every
defense result an upper bound. We built an adaptive harness for the screening benchmark and did not
complete the pre-registered sweep in time for this report; that sweep is the single most valuable
missing measurement, and its absence is why we make no robustness claim about screening beyond the
non-adaptive setting.

\paragraph{Corpus N is a constructed worst case.} All answer-bearing evidence arrives untrusted, and
real deployments will not sit at that extreme. But the arithmetic in \Cref{sec:results-margin}
implies the relationship is linear in the affected fraction rather than thresholded: each question
whose evidence arrives untrusted is lost outright, so a deployment in which 10\% of evidence is
untrusted loses approximately 10\% of its answerable questions. Corpus N therefore fixes the rate of
loss rather than describing an unusual configuration. What we have not done is locate any real
deployment's operating point, and we make no claim about where typical systems sit.

\paragraph{Corpus M is artificial in the opposite direction.} Its untrusted band is drawn entirely
from non-evidence rounds, so suppressing untrusted content removes only distractors. The accuracy
gain to \MixW{} therefore includes a windfall unavailable to any system where untrusted content
carries information. Corpus M bounds the benefit; Corpus N bounds the cost. Together they bracket a
range without locating a point inside it.

\paragraph{The proposed remedy is unevaluated.} We argue from two measured failures that provenance
belongs in retrieval as a bounded occupancy constraint rather than an additive weight. We have not
implemented that gate, and we have not measured whether it degrades gracefully in practice or merely
relocates the failure. The design conclusion is motivated by our results, not demonstrated by them.
It also inherits a dependency we should state rather than leave implicit: an occupancy quota is
still a retrieval-side mechanism keyed on the provenance label, so it addresses the failure mode we
measured --- an additive term with no floor --- without addressing whether the label itself can be
trusted \citep{louck2026tmanm}. A quota and write-time origin binding are complementary, and only
the pair is a defense.

\paragraph{One retriever, one embedding model, one reader.} \Cref{eq:margin} is a statement about
the scoring function, not about any particular embedding space, but whether its limiting case is
reached depends on the similarity distribution a given embedder produces over a given corpus. A
retriever whose in-namespace similarities spread more widely would leave untrusted content some room
to rank. Similarly, the 35\% of utility that survives the undefended arm is partly attributable to
one reader's abstention behaviour; a more credulous reader would retain less, a more sceptical one
more. We do not know how much of either result is model-specific.

\paragraph{Trust labels are assumed correctly assigned.} We study what happens when provenance is
accurate but uninformative about truth. We do not study an adversary who obtains a higher trust
level, an integration that mislabels systematically, or a system in which trust assignment is itself
attackable. \Cref{sec:results-margin} shows that mislabeling in the conservative direction is
already destructive at the corrected weights, which suggests label integrity deserves the same
scrutiny as the ranking function. Concurrent work makes this sharper than a caveat.
\citet{louck2026tmanm} argues that trust-scoring and lineage signals are both malleable, since an
attacker can launder an untrusted origin through the agent's own summarization, a trusted-tool
echo, or manufactured corroboration, and concludes that write-time origin binding is necessary for
any authority decision to be sound. Our results are conditional on a label the adversary cannot
move. That condition is an assumption of this paper, not a property we establish, and the case for
treating it as load-bearing is now stronger than when we made it.

\paragraph{One system.} All measurements come from a single memory implementation. The screening
result generalises by argument rather than by measurement: no content scanner can detect falsity
without knowing the answer, and that argument does not depend on which scanner. The ranking result
generalises to the family of systems that combine provenance and similarity in a weighted sum, by
the arithmetic rather than by replication. Neither has been tested against another implementation,
and systems that treat provenance as a filter, a quota, or a hard constraint are outside the scope
of the argument entirely.

\paragraph{Domain and scale.} LongMemEval is conversational personal-assistant memory. We do not
know how these results transfer to code, clinical, or operational memory, where the base rate of
untrusted content and the cost of abstention both differ. The poisoning arms use a seeded
$n{=}\PoisonN$ subsample against \CleanN{} for the clean baseline; the subsample is drawn once and
reused across all arms, so comparisons are paired, but the absolute accuracies carry the sampling
error of 120 questions.

\paragraph{Measurement caveats we did not resolve.} Latency for the API-backed configurations is not
reported because those runs were served substantially from cache and include rate-limit backoff;
they are not comparable to the locally-executed systems. Write-time screening also refused 109 of
124{,}462 ingested rounds as suspected credential leaks. None were answer-bearing, so no score is
affected, but it is over-defense on ordinary content and we have not characterised its shape.

\section{Conclusion}
\label{sec:conclusion}

We measured what a defended agent memory is worth while it is being poisoned, using the weakest
attack in its class: plainly-worded false statements, generated in one pass, carrying no
instruction and optimized against nothing. At \PoisonFrac{} of the corpus, that attack removes
two-thirds of the memory's value.

Write-time content screening did not engage. The same pipeline that reaches \RecallIndirect{}
recall on indirect injection and flags benign trigger-word text at \FprAegis{} --- an order of
magnitude below the DeBERTa-based detectors we compared against --- refused \ScreenCaught{} of
\PoisonTotal{} poisoned memories. This is not a deficiency of that pipeline. Distinguishing a false assertion from a true one generally
requires external grounding beyond the text being screened. This attack class therefore sits outside
what content-only screening can reliably decide.

That places the defensive burden on the read path, where our results are less comfortable. The
shipped provenance weight was statistically indistinguishable from no defense at all. Raising it
worked, but not for the reason we assumed: at the corrected weight the trust term stops behaving
like a prior and becomes a hard exclusion, because its absolute penalty exceeds half the range
available to similarity. When the untrusted band held only disposable content, exclusion looked like
a defense. When it held the evidence, retrieval collapsed to zero on every one of 120 questions.

For this additive scoring form under the measured similarity regime, the provenance term cannot
express the policy that is wanted.
Any weight large enough to resist an attacker who can shape content is large enough to exclude
untrusted content categorically, because the attacker's achievable similarity advantage and the
corpus's own similarity spread are quantities of the same order. There is no setting of the scalar
that prefers trusted evidence without being willing to discard the only evidence available. We
therefore read these measurements as pointing toward provenance as a bounded occupancy constraint on
retrieved context --- a floor and a ceiling rather than a slope --- and we state clearly that we have
not yet built or evaluated such a mechanism.

Two methodological points generalise beyond this system. First, attack success rate is the wrong
primary metric for memory security: it cannot distinguish a memory that resisted an attack from one
that was rendered useless by it, and it is silent on what a defense costs when nothing is attacking.
Utility retained answers both. Second, a defense evaluated only in the configuration where its
signal is perfectly correlated with the threat has not been evaluated. Our own corrected weight
looked like a success under exactly that condition and revealed a denial-of-service primitive under
one small change of assumption. The mixed-provenance arms cost \$\EvalSpend{} and were the most
informative measurements in this paper.

\paragraph{Future work.} The pre-registered adaptive sweep against the screening pipeline is the
most valuable missing measurement and is next. Beyond it: implementing and evaluating the occupancy
gate; measuring the margin behaviour across embedding models with different similarity spreads to
establish how corpus-dependent \Cref{eq:margin}'s limiting case is; and extending the
utility-under-attack protocol to memory systems outside the additive-scoring family, where the
arithmetic here does not apply and the empirical question is genuinely open.

\bibliographystyle{plainnat}
\bibliography{refs}

\appendix
\section{Reproducibility}
\label{app:repro}

Every table, figure, and prose macro in this source bundle is generated by a single script
(\code{scripts/make_figs.py}) from frozen JSON inputs under \code{data/}. The screening input is a
verbatim copy of \code{benchmarks/injection/results/results.json}. For LongMemEval, the bundle
contains a consolidated snapshot of the primary poisoning measurements documented at the pinned
revision plus the committed mixed-provenance report. Re-running the script regenerates the tables,
figures, and numeric macros without manually editing those outputs.

\paragraph{Artifact.} Source, benchmark harnesses, the attack corpus, and committed aggregate
reports are at:
\begin{quote}
\small
\url{https://github.com/quantifylabs/aegis-memory}\\
revision \code{6d2863083361f7a5c8e12b4512346c94cb453c2c}
\end{quote}
All results in this paper were produced at that revision. The W4.2 primary poisoning values are
also recorded in \code{docs/security/memory-poisoning.md} and are regenerated by
\code{benchmarks/memory/longmemeval/w42_report.py}; the W4.3 mixed-provenance values are recorded
in \code{benchmarks/memory/longmemeval/results/mixed_untrusted_report.json}.

\paragraph{Pinned dataset revisions.} The five screening corpora and their revisions are given in
\Cref{tab:corpora}. The memory benchmark uses LongMemEval\_S at revision \code{2ec2a557}, SHA-256
\code{08d8dad4...7894}, with the synthetic benign corpus produced by generator version
\code{builtin-v1}.

\paragraph{Models.} Screening benchmark: \code{gpt-4o-mini} and
\code{claude-haiku-4-5-20251001} as Stage-4 and LLM-judge backends. Memory benchmark:
\code{claude-sonnet-5} as reader, \code{gpt-4o-2024-08-06} as judge at temperature 0, using
LongMemEval's official judge prompts verbatim. Poison generation:
\code{claude-haiku-4-5-20251001}.

\paragraph{Seeds and sampling.} Seed 42 throughout: bootstrap resampling ($n{=}1000$), the
$n{=}\PoisonN$ question subsample, and the selection of rounds written \texttt{untrusted} in the
mixed-provenance corpora. The subsample is drawn once and reused across every arm, so all
comparisons are paired. Per-round trust assignments are recorded to \code{trust_plan.jsonl}
rather than left recoverable only by replaying the generator.

\paragraph{Determinism.} Model responses are cached under
$(\text{system}, \text{model}, \mathrm{sha256}(\text{prompt}))$, with sampling temperature folded
into the key so that a temperature change produces a fresh cache rather than silently reusing
completions drawn at another setting. Systems whose credentials or model licences were unavailable
are recorded as not-run rather than omitted.

\paragraph{Cost.} The four mixed-provenance arms cost \$\EvalSpend{} in model API usage, measured
rather than estimated and recorded per arm in the released report.

\paragraph{Environment.} Python 3.11.9; \texttt{transformers} 4.53.3, \texttt{torch} 2.12.0 (CPU),
\texttt{datasets} 2.19.1. Latency figures were collected on this configuration and should be read as
relative rather than absolute.

\end{document}